\documentclass{webofc}
\usepackage[varg]{txfonts}   
\begin{document}
\title{Constraining AGN feedback model with SZ profile}
%
%

\author{\firstname{Hideki} \lastname{Tanimura}\inst{1}\fnsep\thanks{\email{hideki.tanimura@ias.u-psud.fr}} \and 
	\firstname{Gary} \lastname{Hinshaw}\inst{2,3,4} \and
	\firstname{Ian G.} \lastname{McCarthy}\inst{5} \and
	\firstname{Ludovic} \lastname{Van Waerbeke}\inst{2} \and
	\firstname{Nabila} \lastname{Aghanim}\inst{1}
	\firstname{Yin-Zhe} \lastname{Ma}\inst{6,7} \and
	\firstname{Alexander} \lastname{Mead}\inst{8} \and
	\firstname{Tilman} \lastname{Tr\"{o}ster}\inst{8} \and
	\firstname{Alireza} \lastname{Hojjati}\inst{2} \and
	\firstname{Bruno} \lastname{Moraes}\inst{9,10}
}

\institute{Universit\'{e} Paris-Saclay, CNRS, Institut d'Astrophysique Spatiale, B\^atiment 121, 91405 Orsay, France \and
           Department of Physics and Astronomy, University of British Columbia, Vancouver, BC V6T 1Z1, Canada \and
           Canadian Institute for Advanced Research, 180 Dundas St W, Toronto, ON M5G 1Z8, Canada \and
           Canada Research Chair in Observational Cosmology \and
           Astrophysics Research Institute, Liverpool John Moores University, Liverpool, L3 5RF, United Kingdom \and 
           Astrophysics and Cosmology Research Unit, School of Chemistry and Physics University of KwaZulu-Natal, Durban, South Africa \and
           NAOC-UKZN Computational Astrophysics Center (NUCAC), University of KwaZulu-Natal, Durban, 4000 \and
           Institute for Astronomy, University of Edinburgh, Royal Observatory, Blackford Hill, Edinburgh, EH9 3HJ, UK \and
           Department of Physics and Astronomy, University College London, Gower Street, London, WC1E 6BT, UK \and
           Instituto de Fisica, Universidade Federal do Rio de Janeiro, 21941-972, Rio de Janeiro, RJ, Brazil
          }

\abstract{Relativistic jets from AGN have a wide range of impacts on galaxy groups and clusters and are key for understanding their formation and physical properties. However, this non-gravitational process is not well understood. Galaxy groups with shallow gravitational potentials are ideal laboratories to study and constrain the AGN feedback model.
We studied hot gas in $\sim$66,000 SDSS LRG halos with an average halo mass of $3 \times 10^{13}$ M$_{\odot}$ using the {\it Planck} tSZ map. We have detected their average tSZ radial profile at $\sim$17$\sigma$ and compared it with the cosmo-OWLS cosmological hydrodynamical simulations with different AGN feedback models. The best agreement has been obtained for the AGN 8.0 model in the simulations. We have also compared our measured tSZ profile with the prediction from the universal pressure profile assuming the self-similar relation and found them consistent if the model accounts for the clustering of neighboring haloes via a two-halo term. 
}
\maketitle
%



\section{Introduction}
\label{intro}

A power-law relation between the electron (or gas) pressure and mass, $P - M^{2/3}$, is valid under the assumption that the galaxy-formation process is dominated by gravity, called self-similar relation. Any deviation from this relation points to the presence of more complex processes such as baryonic feedback effects. X-ray and SZ observations demonstrate that massive galaxy clusters almost follow the self-similar scaling relation \cite{Planck2013b}. On the other hand, numerous X-ray observations find a deficit of baryonic gas inside low-mass halos compared to the cosmic average baryon fraction \cite{Gastaldello2007, Pratt2009, Sun2009, Gonzalez2013}, implying the deviation from the self-similar scaling relation.
The gas deficit may be caused by AGN feedback: the response of the intracluster medium (ICM) to the relativistic plasma ejected by the active galactic nuclei (AGN) \cite{Gitti2012}.
The AGN feedback is key for understanding the evolution of galaxies and the formation of large-scale structures; however, not well understood.

\section{Stacking $y$ map centered on LRGs}
\label{sec:stacking}

\begin{figure}[h]
\begin{center}
\includegraphics[width=0.49\linewidth]{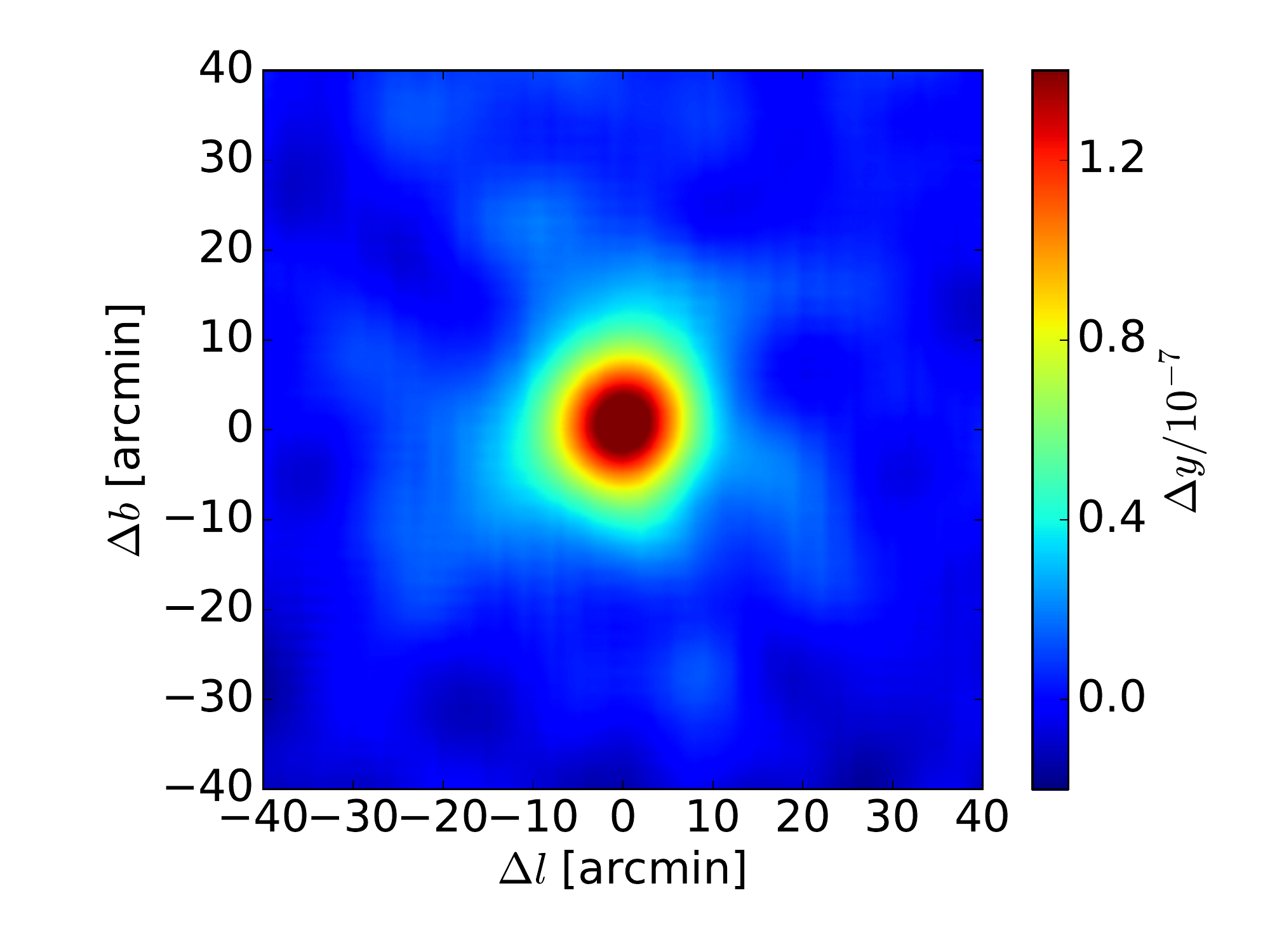}
\includegraphics[width=0.49\linewidth]{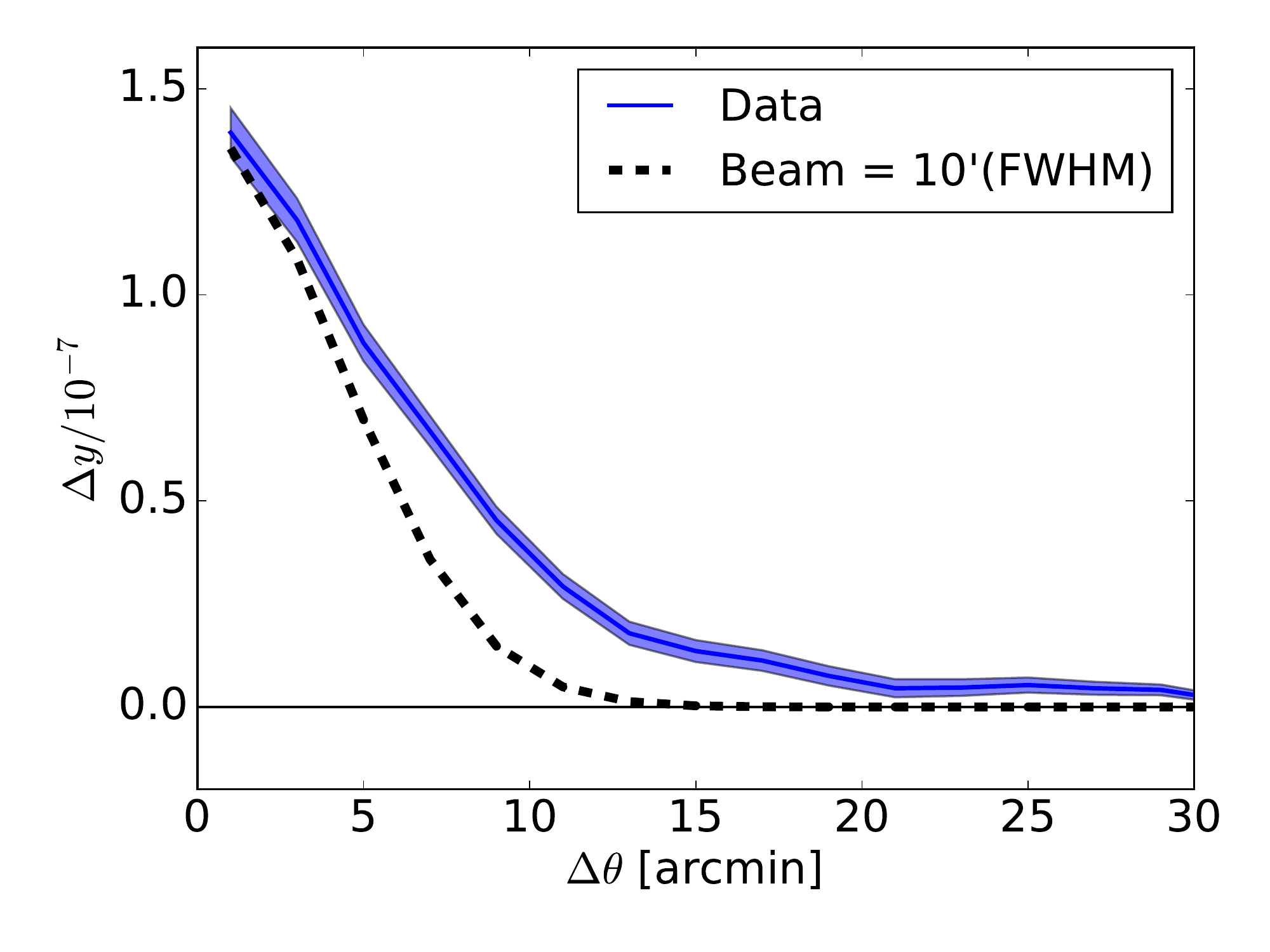}
\end{center}
\caption{
{\it Left}: Average $y$ map centered on 66,479 LRGs. {\it Right}: Average $y$ profile around the LRGs. The 1$\sigma$ statistical uncertainty is represented via the width of the blue line. The $\mathrm{FWHM}=10$ arcmin Gaussian beam of the {\it Planck} $y$ map is shown as a black dashed line for comparison, the peak of which is normalized to the center of the LRGs' $y$ profile. } 
\label{fig:stacking}
\end{figure}

We performed a stacking analysis using the {\it Planck} $y$ map \cite{Planck2016-XXII} for Sloan Digital Sky Survey (SDSS) data release 7 (DR7) luminous red galaxies (LRGs) at $0.16 < z < 0.47$ \cite{Blanton2005}, which are considered to be mostly central galaxies in dark matter halos \cite{Reid2009}. We constructed an average $y$ profile centered on the 66,479 LRGs with the stellar mass\footnote{The stellar masses of the LRGs are provided in the New York University Value-Added catalogue (NYU-VAGC) at http://sdss.physics.nyu.edu/vagc/} of $10^{11.2} \leq M_{*}/M_{\odot} \leq 10^{11.7}$, roughly corresponding to halo masses of $10^{13}\leq M_{500}/M_{\odot} \leq 10^{14}$. We detected a significant signal at $\sim 17 \sigma$ out to $\sim$30 arcmins well beyond the extent of the 10 arcmin beam of the {\it Planck} $y$ map in Fig.~\ref{fig:stacking}.

\section{Comparison with the cosmo-OWLS hydrodynamic simulations}
\label{sec:sims}

\begin{figure}[h]
\begin{center}
\begin{minipage}{0.49\linewidth}
\includegraphics[width=\linewidth]{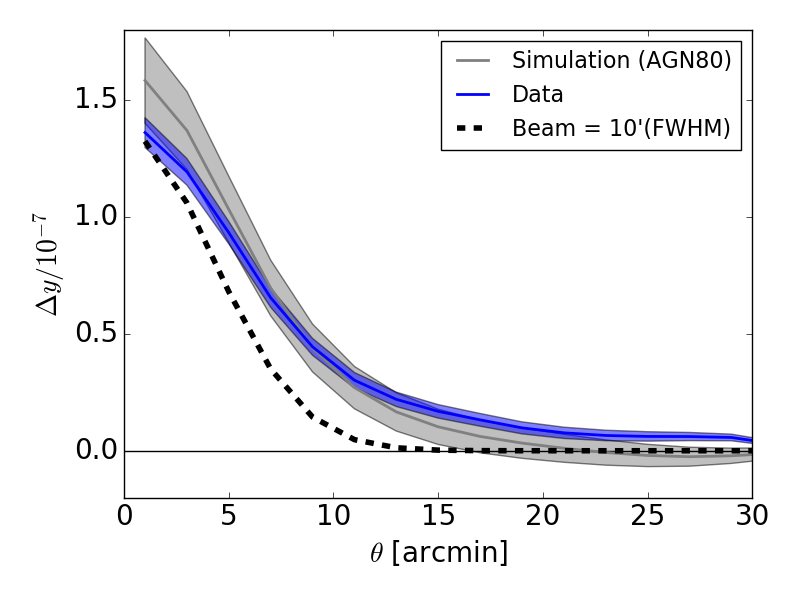}
\end{minipage}
\begin{minipage}{0.49\linewidth}
\includegraphics[width=\linewidth]{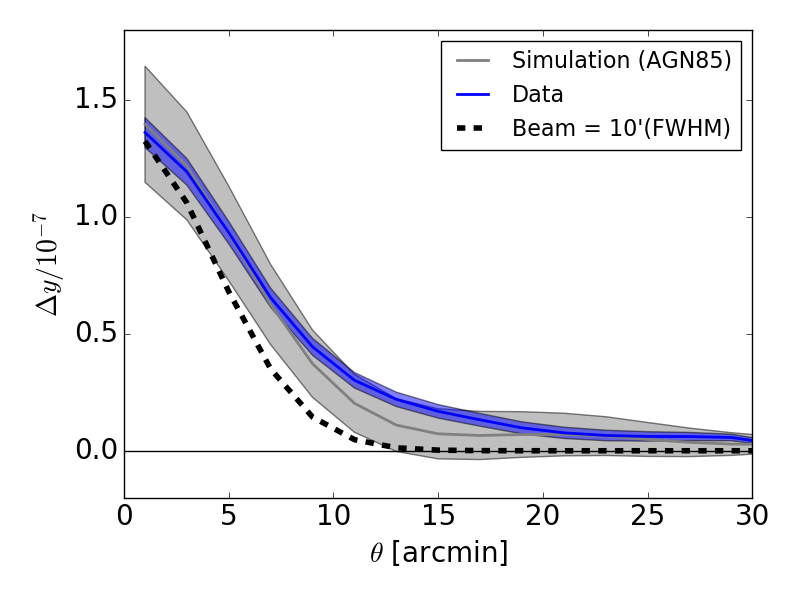}
\end{minipage}
\begin{minipage}{0.49\linewidth}
\includegraphics[width=\linewidth]{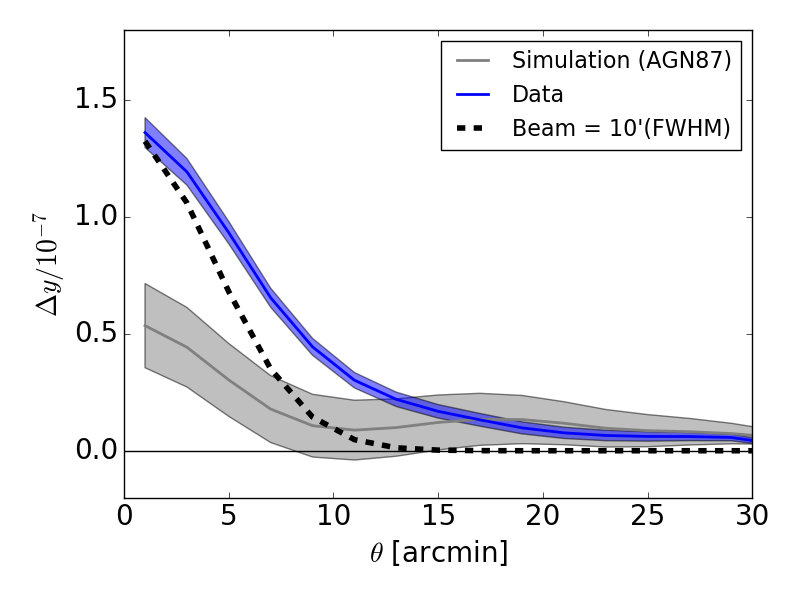}
\end{minipage}
\begin{minipage}{0.49\linewidth}
\includegraphics[width=\linewidth]{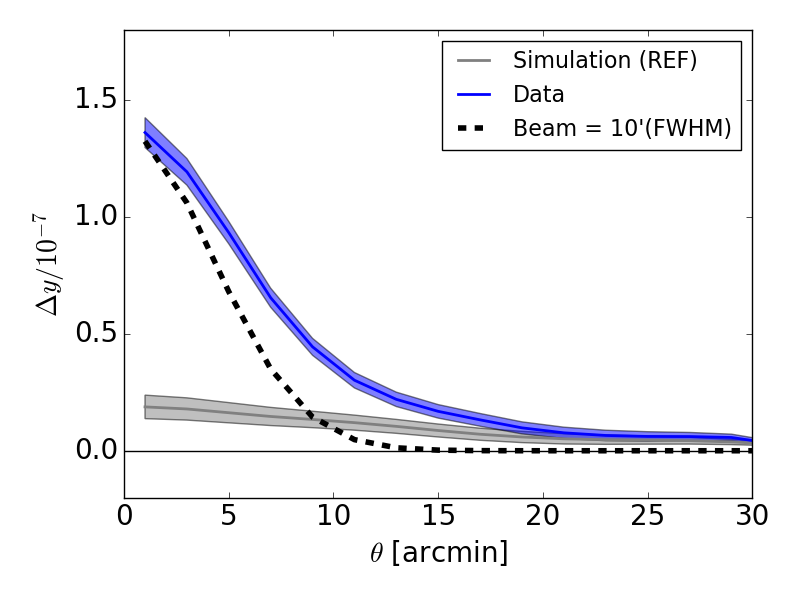}
\end{minipage}
\end{center}
\caption{The average $y$ profile around LRGs (blue) is compared to the $y$ profiles of the simulated central galaxies (gray) in different AGN feedback models respectively. In each case we have matched the stellar mass and redshift distributions from the simulations to be the same as those in the data. {\it Top left}: AGN 8.0 model, {\it Top right}: AGN 8.5 model, {\it Bottom left}: AGN 8.7 model, and {\it Bottom right}: REF model.  }
\label{fig:comp}
\end{figure}

We compared the average $y$ profile around LRGs with predictions from the cosmo-OWLS cosmological hydrodynamical simulations \cite{Brun2014, Daalen2014, McCarthy2014}.
The simulations are designed with cluster cosmology and large-scale structure surveys in mind. They consist of box-periodic hydrodynamical simulations and have volumes of 400$h^{-1} \rm{Mpc}^3$ with $1024^3$ baryonic and dark matter particles.
They simulated five different models (Table.~\ref{tab:sims}), which are compared with our measured $y$ profile in Fig.~\ref{fig:comp}.

In the comparison, a clear difference between the data and REF model can be seen. 
In general, energy released from the center of a halo heats cluster gas, this in turn prevents cooling and thus the star formation around the central region. Therefore, if we consider halos of the same total mass, the stellar mass of the central galaxy is decreased as the power of the central AGN is increased. Since we select central galaxies based on stellar mass, lower-mass halos are selected in the REF model compared to the models that include AGN feedback. This is apparent as the lower central peak value of the simulated $y$ profiles in the REF model compared to the AGN models. We also see a visible trend in that the higher the power of AGN feedback, the lower the peak of $y$ profile. This is due to the fact that the AGN feedback ejects gas from the center of halos outward and the overall gas density is lowered.

In summary, this comparison agrees best with simulations that include AGN feedback (AGN 8.0) but not with simulations that do not include it (REF) or with simulations with very violent AGN feedback (AGN 8.7). The reduced $\chi^2$ values are 1.0 (AGN 8.0), 1.3 (AGN 8.5), 9.9 (AGN 8.7), and 75.9 (REF), respectively. This result is consistent with other studies showing that the AGN 8.0 model reproduces a variety of observed gas features in optical and X-ray data (e.g., \cite{Brun2014}).

\begin{table}[h]
\centering
\caption{The baryon feedback models in the cosmo-OWLS simulations.}
\begin{tabular}{|l|l|l|l|l|l|} \hline
Simulation & Cooling & Star formation & SN feedback & AGN feedback & $\Delta T_{\rm heat}$ \\ \hline
NOCOOL & No & No & No & No & ... \\ 
REF & Yes & Yes & Yes & No & ... \\ 
AGN 8.0 & Yes & Yes & Yes & Yes & $10^{8.0}$ K \\ 
AGN 8.5 & Yes & Yes & Yes & Yes & $10^{8.5}$ K \\
AGN 8.7 & Yes & Yes & Yes & Yes & $10^{8.7}$ K \\ \hline
\end{tabular}
\label{tab:sims}
\end{table}

\section{Halo model prediction with the UPP}
\label{sec:pred}

\begin{figure}[h]
\begin{center}
\includegraphics[width=0.5\linewidth]{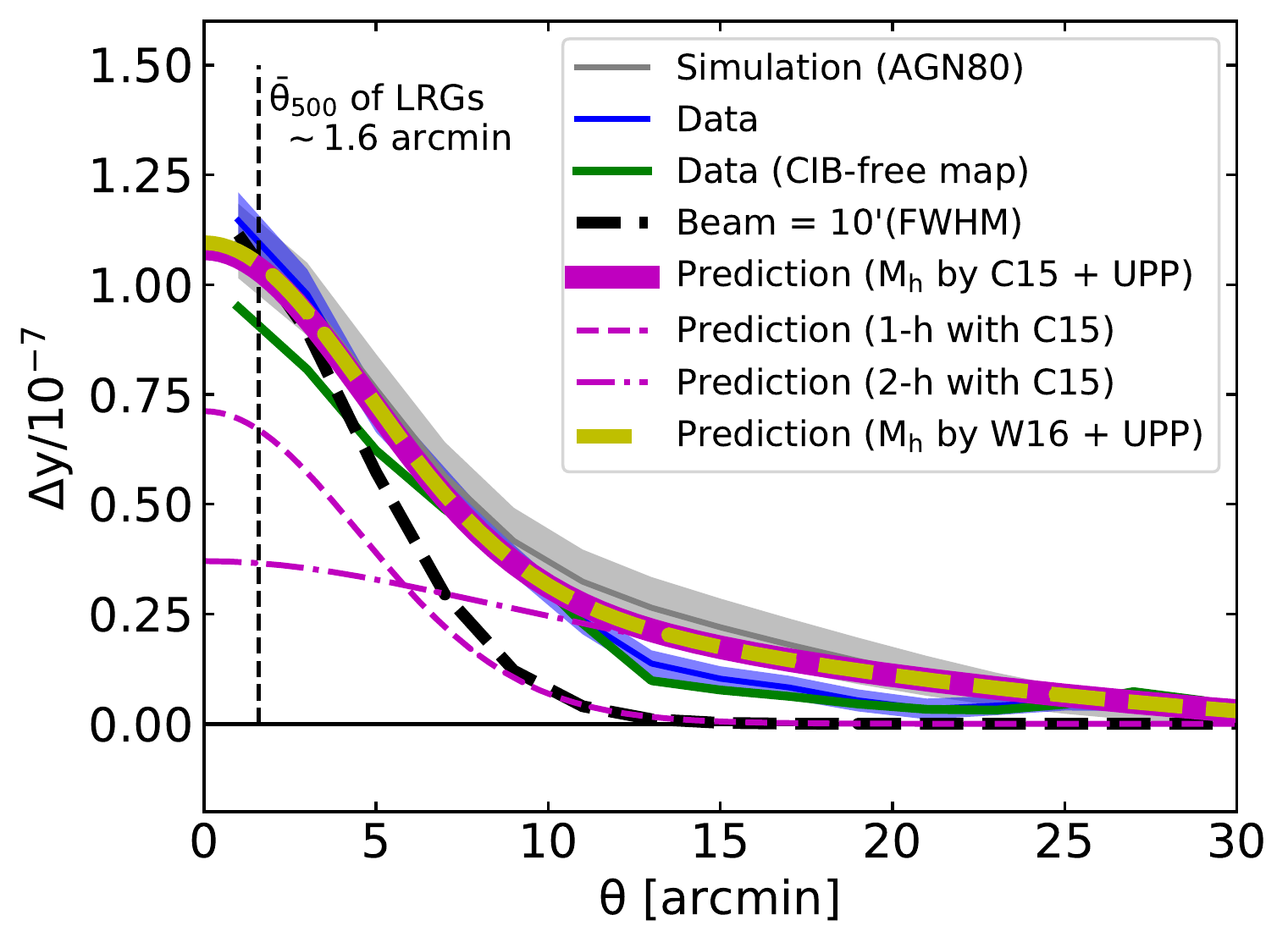}
\caption{The average $y$ profile around 66,479 LRGs (blue) is compared to the predictions using a halo model with the halo mass function and halo bias \cite{Tinker2010} and UPP \cite{Planck2013-V}. The halo masses of the LRGs are estimated using either the SHM relation of \cite{Coupon2015} (C15-SHM: magenta) and \cite{Wang2016} (W16-SHM: yellow). The one-halo (dashed line in magenta) and two-halo (dash-dotted line in magenta) terms are shown separately for the model prediction using the C15-SHM. The $y$ profile of the simulated central galaxies in the AGN 8.0 simulation is shown in grey.  To show an impact of beam, the average angular size of the LRGs ($\bar{\theta}_{500} \sim 1.6$ arcmin) is shown in vertical black dashed line.} 
\label{fig:prediction}
\end{center}
\end{figure}

The average $y$ profile around the LRGs is also compared with a prediction using the halo model with the halo mass function and halo bias \cite{Tinker2010} and a universal pressure profile (UPP) \cite{Planck2013-V} in Fig.~\ref{fig:prediction}. 
The stellar masses of the LRGs are converted to halo masses using the stellar-halo mass relation from either \cite{Coupon2015} (C15-SHM) or \cite{Wang2016} (W16-SHM), estimated using gravitational lensing measurements.
Then, the UPP is scaled to different masses of halos, assuming a self-similar relation. The predicted $y$ profile is consistent with the data, but only if we account for the two-halo clustering term in the model and if we assume the stellar-halo mass relation from C15-SHM or W16-SHM. This consistency may imply that the UPP, estimated for massive galaxy clusters in the mass range of $10^{14.4} - 10^{15.3}$ M$_{\odot}$, can be applied even in low-mass halos down to $\sim 10^{13.5}$ M$_{\odot}$ in a self-similar manner.

\section{Conclusion}
\label{sec:con}
According to the study with the cosmo-OWLS simulations \cite{Brun2015}, the AGN 8.0 simulation predicts more extended pressure profiles around low-mass halos than the UPP. However, the AGN 8.0 model also reproduces our observed $y$ profile, supporting a self-similar assumption. This apparent inconsistency is explained by the fact that the deviations from the self-similar relation in the AGN 8.0 simulation are mainly confined to halo masses below $M_{500} \sim 10^{13.5}$ M$_{\odot}$, which roughly corresponds to the average mass of our sample. Furthermore, the impact of coarse angular resolution of the {\it Planck} $y$ map is not negligible in our analysis: the UPP and AGN 8.0 pressure distributions only differ significantly on scales of $r < r_{500}$ in the simulations, in which they are compared without the beam. Thus, data from higher-resolution tSZ maps (such as those from ACT or SPT with FWHM of order an arcminute) would be necessary in this regard.

\section*{Acknowledgement}
{\small This research has been supported by the funding for the ByoPiC project from the European Research Council (ERC) under the European Union's Horizon 2020 research and innovation programme grant agreement ERC-2015-AdG 695561.}

%
%

\end{document}